\documentclass[10pt,journal]{IEEEtran}
\usepackage[numbers,square]{natbib}
\bibpunct{[}{]}{;}{a}{}{,}
\usepackage[pdftex]{graphicx,hyperref}
\usepackage{txfonts}
\usepackage{subfigure}

\begin{document}

\title{The non-coplanar baselines effect in radio interferometry: The W-projection algorithm}
\author{T.~J.~Cornwell,  K.~Golap and S.~Bhatnagar
\thanks{T.~J.~Cornwell is with the Australia Telescope National Facility, Epping, NSW,
  Australia} 
\thanks {K.~Golap and  S.~Bhatnagar are with the National Radio Astronomy Observatory, Socorro,
NM 87801, USA}
\thanks{Associated Universities
Inc. operates the National Radio Astronomy Observatory under
cooperative agreement with the National Science Foundation} 
}

\maketitle
\begin{abstract}We consider a troublesome form of non-isoplanatism in 
synthesis radio telescopes: non-coplanar baselines. We present a novel
interpretation of the non-coplanar baselines effect as being due to
differential Fresnel diffraction in the neighborhood of the array
antennas.

We have developed a new algorithm to deal with this effect. Our new
algorithm, which we call ``W-projection'', has markedly superior
performance compared to existing algorithms. At roughly equivalent
levels of accuracy, W-projection can be up to an order of magnitude
faster than the corresponding facet-based algorithms.  Furthermore,
the precision of result is not tightly coupled to computing time. 

W-projection has important consequences for the design and operation
of the new generation of radio telescopes operating at centimeter and
longer wavelengths.

\end{abstract}

\date{Received date; accepted date}
\markboth{W-projection}
\maketitle

\newcommand{\aipspp}{{\sl AIPS++/CASA}}
\newcommand{\ncel}{\sqrt{1-\ell^2-m^2}}
\newcommand{\x}{{\bf{x}}}
\newcommand{\xp}{{{\bf x}_p}}
\newcommand{\xpi}{{{\bf x}_p^i}}
\newcommand{\bu}{{\bf{u}}}
\newcommand{\DR}{{\bf DR}}
\newcommand{\FI}{{\bf FI}}
\newcommand{\VSNR}{{\bf VSNR}}
\newcommand{\SNR}{{\bf SNR}}
\newcommand{\aap}{{Astron. \& Astrophys.}}


\section{Introduction}

Wide-field imaging with synthesis radio telescopes can be limited
by a number of effects:

\begin{itemize}
\item The intrinsic performance of the deconvolution algorithms.
\item Time and frequency averaging.
\item The finite size of the primary beam.
\item The individual, angular, frequency, polarization, and temporal 
variations in the antenna primary beams.
\item Non-isoplanatic atmospheres.
\item The non-coplanar baselines effect.
\end{itemize}

Effective and reasonably efficient algorithms exist to correct for
many of these effects \citep[see][Chapter 19]{Tayloretal1999}.

Early interferometric arrays were often designed to be aligned
East-West, in which case the baselines are coplanar and no error
arises (for an excellent review of the development of radio
interferometry see chapter 1 in the standard text by Thompson, Moran,
and Swenson \citep[][]{Thompsonetal2001}). Early {\em non-coplanar}
arrays had limited sensitivity and so only the brightest sources at
the centre of a small field of view were of interest. As the
sensitivity improved over the years with lower system temparatures, it
became necessary to find and remove the effects of the other bright
sources in the antenna primary beam.

Only as non-coplanar arrays become sufficiently sensitive did the need
for a full field of view correction algorithm become apparent.  For
the Very Large Array (VLA), this transition occured with the commissioning
of the 327MHz observing system in the early nineties. At that point,
Cornwell and Perley developed a faceted algorithm
\citep{CornwellPerley1992} which has been used in one form or another
since then. Reaching the thermal noise sensitivity limit on the VLA
required using the faceted algorithm to correct for the sidelobes from
the myriad background sources in any field. However, the faceted
algorithm is typically 100 to 1000 times slower than the simple two
dimensional inversion, and so the need for a faster algorithm
remained.

In this paper, we briefly re-examine various algorithms for dealing
with the non-coplanar baselines effect before presenting our new
algorithm, its implementation, performance and implications for the
new generation of radio telescopes.


\section{Overview of the non-coplanar baselines problem}

Imaging in Radio Astronomy is determining the brightness distribution,
$I$, at a given frequency as a function of some angular coordinates,
$(\ell,m)$, on the sky . The preferred radio interferometric
coordinates are direction cosines \citep[][]{Thompsonetal2001}. The
output of the correlation of a pair of antennas (interferometer
elements) is referred to as the visibility.

An interferometer measures the spatial coherence function of the
electric field between two points at positions  $\overrightarrow{r_i}$  and $\overrightarrow{r_j}$:

\begin{equation}
\label{EQN:COHERENCE}
V_{i,j}  =  \left< E(\overrightarrow{r_i},t)E^*(\overrightarrow{r_j},t)\right>_t
\end{equation}

The response of a narrow-band phase-tracking interferometer to
spatially incoherent radiation from the far field can be expressed by
the following relation between the spatial coherence, or visibility, $V(u,v,w)$, and the spectral intensity, or brightness, $I(\ell,m)$. The values $u,v,w$ are the  components of the vector between the two
interferometer elements ($\overrightarrow{r_j}-\overrightarrow{r_i}$) expressed in units of wavelength of the
radiation (see \citep[][Chapter 3]{Thompsonetal2001}).

\begin{equation}
\label{EQN:FULL}
V(u,v,w)  =  \int{I(\ell,m)\over\ncel} e^{-2\pi i[u\ell + vm + w(\ncel-1)]}d\ell dm
\end{equation}

When the magnitude of the term $2 \pi w(\ncel-1)$ is much less than
unity, it may be ignored, and a two-dimensional Fourier relationship
results. The recovery of the brightness distribution, $I(\ell,m)$,
involves a Fourier transform and a deconvolution process.  A
deconvolution is needed as $V(u,v,w=0)$ is sampled discretely on the
$u,v$ plane and thus the Fourier transform of that sampling function
(known as Point Spread Function or PSF) needs to be removed
iteratively (see appendix A) . The image made from the Fourier
transform of the sampled visibilities is referred to as the {\em dirty
  image}.

When the $2 \pi w(\ncel-1)$ term is comparable to or exceeds unity, a
two-dimensional Fourier transform cannot be used. As a consequence, it
is not possible to estimate the sky brightness by simple Fourier
inversion of the measured visibility.

With the assumption that the maximum $w \approx \frac{B}{\lambda}$,
the value of the extra phase term is roughly:

\begin{equation}
{B \lambda \over D^2} = \left(r_F\over D\right)^2
\end{equation}

where $B$ is the maximum baseline length  , $D$ is the antenna diameter,
and $\lambda$ is the observing wavelength. The parameter $r_F$ is the
Fresnel zone diameter for a distance $B$. The role of the Fresnel zone
diameter is somewhat curious; we have an explanation below. 

It is useful to work with the inverse:

\begin{equation}
N_F = {D^2\over B \lambda}
\end{equation}

Wide-field imaging is affected by this {\em non-coplanar baselines}
effect when the Fresnel number $N_F$ is less than unity: this occurs for
small apertures, long baselines, or long wavelengths. In optics
terminology, the effect is a {\em vignetting}: a limitation of the
field of view due to the optical system.

A crucial distinction must be made between interferometer space and
Fourier space. A single interferometer does not measure a single
Fourier component (unless $w=0$). We therefore use the term uvw-space
to denote the space in which the measurements are made. 

\section{Review of existing algorithms for non-coplanar baselines}

A number of algorithms for dealing with non-coplanar baselines have
been proposed. The best of these algorithms have been scientifically
successful. However, for new telescopes such as the Expanded Very
Large Array \citep{Perley2000}, and the Square Kilometer Array
\citep{Ekers2003}, the extra computational load is predicted to be
large. In the course of the development of our new algorithm, we
reviewed all of those listed below to see if any substantial increase
in performance could be made. Cornwell and Perley \citep{CornwellPerley1992} 
described many of these algorithms. We repeat their description here with
some additions based on improved understanding.


\subsection {Fourier sum:}  

Equation (\ref{EQN:FULL}) may be numerically 
integrated using a pixellated
image. This can be arbitrarily accurate but is nearly always
prohibitively expensive. A useful compromise is to perform Fourier
sums for the bright pixels and some other more approximate transform
for the other pixels.

\subsection {Component models:}  

The sky brightness can be modeled by a collection of discrete
components, drawn from a fixed repertoire of component types, the
Fourier transform of which may be calculated analytically.  In the
context of wide-field imaging, this approach allows relaxation of some
of the tolerances on, for example, number of facets (see \ref{SEC:FACETS} below), since the
brightest emission is modeled by exactly transformable components.

\subsection {Warped snapshots:}

For instantaneously planar sampling, $w$ can be eliminated \citep[see][]{Bracewell1983, Brouw1969} from
equation (\ref{EQN:FULL}).  If $Z$ is
the Zenith angle, and $\chi$ is the parallactic angle at the time of
observation, then the relationship between sky brightness and
visibility can then be expressed as:

\begin{equation}
\label{EQN:PLANAR}
V(u,v,w) = \int{I(\ell,m)\over\ncel}
   e^{-2\pi i[u\ell' + vm']}d\ell dm
\end{equation}
where \citep[][Chapter 14]{Perleyetal1989}:

\begin{eqnarray}
\label{EQN:NEWCOORDS}
\ell' & = & \ell + \tan(Z) \sin(\chi) \left(\ncel - 1\right)\\
m' & = & m - \tan(Z) \cos(\chi) \left(\ncel - 1\right)
\end{eqnarray}

Thus a two dimensional Fourier transform between sky brightness and
visibility holds at any instant but at the cost of a coordinate
distortion in the sky plane. This can not be corrected via a simple
linear coordinate transform in uvw-space and so image plane regridding
of each snapshot, either before or after deconvolution, is required.
Unfortunately, the required high precision image plane coordinate
transform is computationally expensive.

It remains a  fact that none of the algorithms 
currently in use (including our new algorithm discussed below)
actually make any use of the instantaneous planarity of the array.

\subsection {3D transforms:}

Equation (\ref{EQN:FULL}) may be embedded in a three dimensional \citep{Clark1973} space with axes
$(\ell,m,n)$.

\begin{equation}
\label{EQN:FULL3D}
V(u,v,w) = \int{I(\ell,m)\delta\left(n-\ncel\right)\over n}
e^{-2\pi i[u\ell + vm + wn]} d\ell dm dn
\end{equation}

This three dimensional Fourier transform may be implemented using an
FFT in all axes, or if the range in $n$ is small, FFTs in $(\ell,m)$
with DFT in $n$. The principal drawback is that for large field of
view, the interior of the cube is largely devoid of true emission.

\subsection {Image-plane facets:}

\label{SEC:FACETS}

Any given widefield  image can be considered a sum of smaller images (or commonly called facets). If each facet size is so chosen that $w$ term in equation \ref{EQN:FULL} is near zero, therefore equation \ref{EQN:FULL} can be transformed to a sum of Fourier transforms \citep{CornwellPerley1992}.   

Assume that the image plane is divided 
into $N_{facets}$ by $N_{facets}$ facets. Each facet is imaged
separately in a minor cycle, and then reconciled in a major
cycle (appendix A describes the concept of major/minor cycles in deconvolving images). Separate images are made for each
facet and then these are reprojected to a common  plane after
deconvolution. Each facet can be  deconvolved separately using the
appropriate PSF. The original Software developement Environment (SDE) {\tt dragon} program \citep{SDE1995} and the Astronomcal Image Processing System (AIPS {\em http://www.aips.nrao.edu})
IMAGR task both use this approach.

The number of facets required along any axis is proportional to
inverse of the Fresnel number. More accurately, the number of facets
needed in $\ell$ and $m$ is:

\begin{equation}
N_{facets} = {\pi \Theta \sigma_W\over\sqrt{32 \delta A}}
\end{equation}

where $\Theta$ is the field of view in radians, $\sigma_w$
is the dispersion in $w$, and $\delta A$ is the maximum tolerable
amplitude loss. If the positions of the sources away from the
phase center are not required to any accuracy, then $\sigma_W$ should
be the residual value after removing a best fitting plane.

As the facet size shrinks all the way down to one pixel, the
image-plane facet algorithms becomes just a Fourier sum. One can
therefore think of the image-plane facet algorithm as combining the
virtues of the direct Fourier sum and FFTs.

Major cycle calculation of the residual images is quite
straightforward but minor cycle deconvolution is more difficult. There
are two problems. First, the facets inevitably overlap on the image
plane. Dealing with this requires complex image plane clean boxes or
complicated logic.  Second, the emission often spans multiple
facets. Deconvolution across the facets is difficult, and instead some
form of feathering adjacent facets is used. A final complication is
that the facetting is in some senses frozen in and cannot easily
be increased as, for example, a deconvolution goes to deeper and
deeper levels.

\subsection {uvw-space facets}

An alternative to image-plane facets is to project the $(u,v)$
coordinates space for each facet onto one tangent plane during the
gridding and Fourier transform steps in imaging (See
\citet{Saultetal1999} for mathematical details).  This is fast
(involving only a matrix multiply of each $(u,v)$) and avoids having
to deal with a large number of facet images. Since the residual image
is contiguous in the image plane, deconvolution may encompass the
entire image. This algorithm is available in \aipspp ({\em http://casa.nrao.edu}).


It is generally believed that either image- or uvw- faceting is needed
to account for the shift-variant nature of the point spread
function. In fact, the PSF varies at a level comparable to or below
the approximations inherent in assuming decoupling of the facets
during the minor cycle. This means that there is no real advantage in
using the local, rather than average, PSF in a major/minor cycle
algorithm \citep{Clark1980}. This realization opens up the possibility of
using algorithms that calculate one residual image for the entire
field. The uvw-space facets algorithm can be viewed as doing this, of
course, but there are also other approaches possible. In the next
section, we discuss an algorithm that does generate one residual image
for the entire field.

\section{W reprojection}

Since the problem largely originates with the $w$ part of $(u,v,w)$,
it is worth asking if there is any way to project $w$ out of the
problem, thus allowing a two dimensional Fourier transform to a single
image to be used. \citet{FraterDocherty1980} noted that projection
from a single plane $w$ to $w=0$ is possible, and they proposed using
Clean to solve the resulting convolution relationship. Frater and
Docherty consider only the (unusual) case of all measurements
occurring on a single plane. The novel part of our contribution is to
realize that their equation allows reprojection to and from any
position in $(u,v,w)$ space from and to the $w=0$ plane by convolution
with a known kernel. To derive this result, we must rewrite equation
(\ref{EQN:FULL}) as a convolution between the Fourier transform of the
sky brightness and the Fourier transform of an image plane phase term
parametrized by $w$.

\begin{equation}
\label{EQN:WGCF}
V(u,v,w) = \int{I(\ell,m)\over \ncel} G(\ell,m,w) \ e^{-2\pi i[u\ell + 
vm]}  d\ell dm 
\end{equation}

\begin{equation}
\label{EQN:GCF}
G(\ell,m,w) = e^{-2\pi i[w(\ncel-1)]}
\end{equation}

Applying the Fourier convolution theorem, we find that:

\begin{equation}
\label{EQN:WGCFFT}
V(u,v,w)=\tilde{G}(u,v,w) * V(u,v,w=0)
\end{equation}

where $\tilde{G}(u,v,w)$ is the Fourier transform of $G(u,v,w)$

To understand the form of $\tilde{G}(u,v,w)$, we can use a small angle
approximation:

\begin{equation}
\label{EQN:GCFA}
G(\ell,m,w) = e^{\pi i[w(l^2+m^2)]}
\end{equation}

\begin{equation}
\label{EQN:GCFUVW}
\tilde{G}(u,v,w) = {i\over w} e^{-\pi i[{(u^2+v^2)\over w}]}
\end{equation}

While mathematically trivial, this has remarkable algorithmic
implications for radio interferometry: the visibility for non-zero $w$
can be calculated from the visibility for zero $w$ by convolution with
the known function $\tilde{G}(u,v,w)$. Thus the three-dimensional
function $V(u,v,w)$ is determined from the two-dimensional function
$V(u,v,w=0)$.  This holographic result is due to the fact that the
original brightness is confined to a two-dimensional surface (the
celestial sphere).

Since the non-coplanar baselines effect is both crucial to wide-field
radio synthesis imaging and somewhat difficult to understand, we will
take some care in interpreting this relationship. The geometry is
shown in Figure 1. Consider radiation propagating from a small range
of angles centered around the vertical axis. If the electric field
time sequences at the points A and B were to be correlated then the
correlation would be the two-dimensional Fourier transform of the sky
brightness. However, the correlation is actually between the electric
field at points A and B'. On propagating from B to B', the electric
field inevitably diffracts. It is this diffraction that prevents the
correlation AB' from being the same as correlation AB. It is easier to
understand this by appealing to reciprocity and considering instead
the transmission case where electric fields are emitted from A and
B'. The electric field on the plane AB is a diffracted version of the
electric field at B'. As the antenna diameter increases, the effect of
diffraction clearly decreases.

We can turn this analysis of the physics into an equation for the
correction. Suppose that point A is at the origin of the $(x,y,z)$
space, B is at $(x=x_a, y=y_a, z=0)$, and that B' is at
$(x=x_a,y=y_a,z=w\lambda)$. The correlation is defined as:
\begin{equation}
\left< \Psi(0,0,0) \Psi^{*}(x_a,y_a,w\lambda) \right>
\end{equation}

To calculate this correlation, we need to be able to relate the
electric fields on the two planes.  Since the region around the
antennas is source-free, the electric field $\Psi$ on plane AB may be
propagated to the plane A'B' using diffraction theory.  If the
distance between the two planes is sufficiently small, Fresnel
diffraction theory must be used \citep{Goodman2002}.

\begin{equation}
\label{EQN:FRESNEL}
\Psi(x,y,z=w\lambda)= {e^{-2\pi i w}\over i \lambda^2 w} \int \Psi(x',y',z'=0)e^{{{\pi i}\over{\lambda^2 w}}((x-x')^2+(y-y')^2)} dx'dy'
\end{equation}

Re-expressing in terms of $(u,v)$, we reproduce the convolution
relationship (equation \ref{EQN:WGCFFT}) between the visibility $V(u,v,w)$ and
$V(u,v,w=0)$.

The physical cause of the convolution relationship is therefore the
Fresnel diffraction of the electric field on propagating from plane AB
to plane A'B'. The size of the diffraction pattern in wavelengths is
given by $r_F/\lambda \sim \sqrt{w}$.

\begin{center}
\begin{figure}[htb]
\label{FIG:DIFFRACTION}
\begin{center}
\includegraphics[width=9cm]{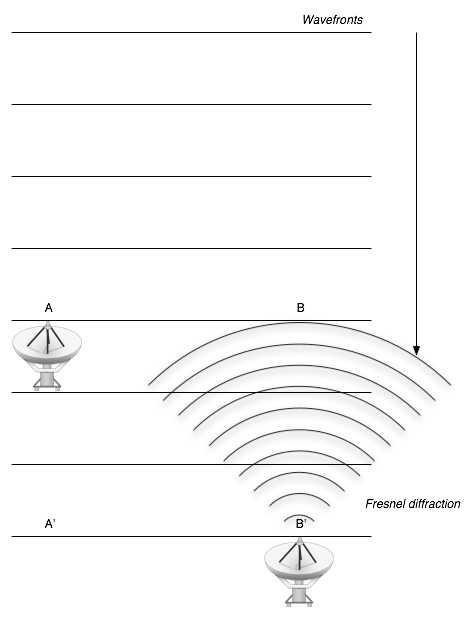}
\end{center}

\caption{Diffraction of electric field on propagation from plane AB to
  plane A'B'. The correlation between the electric fields at point A
  and B is a two-dimensional Fourier transform of the sky
  brightness. Radiation at B is diffracted on propagation to plane
  A'B', and so the measured correlation is no longer a Fourier
  transform of the sky brightness. Alternatively, in transmission, the
  electric field transmitted at B' undergoes the Fresnel diffraction
  on propagation to the AB plane and so the correlation structure of
  the emitted radiation is altered. By the reciprocity theorem, these
  two descriptions are physically equivalent.}
\end{figure} 
\end{center}

In figure (2), we show schematically the effect of this diffractive
projection in $(u,v,w)$ space. A sample taken at $(u,v,w)$ is spread
by diffraction over the $(u,v,w=0)$ plane. This may be interpreted as
the sensitivity of a single sample point at non zero $w$ to a range of
spatial frequencies across the field of view. From the sampling
perspective, the effective diameter of the antennas in an
interferometer is roughly the size of the Fresnel zone $r_F$. This can
be very large - for the A configuration of the VLA at 74 MHz, this is
about $r_F \sim 350m$.

\begin{center}
\begin{figure*}[htb]
\label{FIG:WPROJECT}
\begin{center}
\includegraphics[width=15cm, height=10cm]{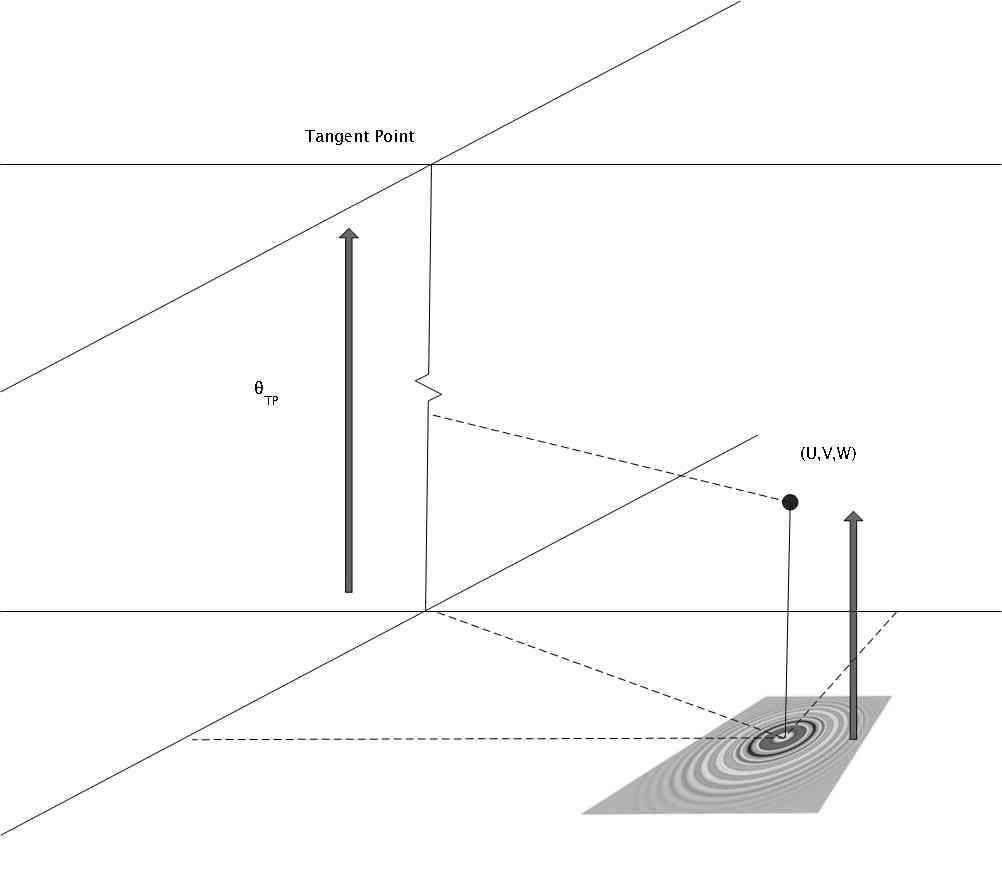}
\end{center}

\caption{Schematic of the projection of a single $(u,v,w)$ sample onto the
$(u,v,w=0)$ plane.}
\end{figure*}
\end{center}

In summary, interferometers with the same $(u,v)$ but different $w$
provide substantially different information on the sky
brightness. Hence, an interferometer can only be said to measure a
{\em single} Fourier component of the sky brightness if $w=0$. For
non-zero $w$, an interferometer is not a device for measuring a single
Fourier component. In principle, then, by measuring at a fixed $(u,v)$
for a range of values of $w$, one can recover information on Fourier
components within $r_F/\lambda$ of the nominal Fourier component
sampling $(u,v)$. This is similar to the principle in mosaicing
\citep{Cornwell1988} where by changing the pointing direction of the
antennas in an interferometer, one can recover Fourier components
within $D/\lambda$ of the nominal. However, in this case, one does not
easily have access to samples spread along the w axis and so
w-synthesis is less useful than mosaicing.

\section{Practical details of an algorithm for W-projection}

\subsection{General structure} 

Given a model of the sky brightness, we can predict the visibility on the
$(u,v,w=0)$ plane by a two-dimensional Fourier transform in $(\ell,m)$. Using the
convolution function project this from the $(u,v,w=0)$ plane to any specific point
$(u,v,w)$. The calculation in this direction (image to uvw-space) is limited only by
numerical errors. Going in the opposite direction (uvw-space to an image) is more
difficult. There is no inverse transform and so we have to rely upon iterative
algorithms. As shown in Appendix A, we need to be able to apply the transpose of the
image to uvw-space operation. In detail, this proceeds as follows: we project each
$(u,v,w)$ point onto the $(u,v,w=0)$ plane using the W-projection function. We then
Fourier transform (using a two-dimensional transform) these gridded visibilities to
the image plane, where we have thus obtained a dirty (or residual) image on the
tangent plane. The dirty image itself can be a good estimate of the sky brightness,
depending upon the sidelobe level of the synthesized beam. More usually, though,
deconvolution will be required. In this situation, the residual image may be used in
a minor cycle deconvolution algorithm to update the model.

In the minor cycle, we use a single PSF calculated for the center of the field. This
introduces errors comparable to the typical sidelobes so we terminate the minor
cycle before these errors become significant.

\subsection{Aliasing}

To limit aliasing, one must apply an additional tapering function
in the image plane. The effective convolution function then becomes
the Fourier transform of:

\begin{equation}
\label{EQN:GCFT}
G_T(\ell,m,w) = T(\ell,m) e^{-2\pi i[w(\ncel-1)]}
\end{equation}

Therefore to evaluate the visibility predicted for a pixellated model
$I$ of the sky brightness, we do the following:

\begin{enumerate}
\item Multiply the sky brightness model $I$ by the taper
function $T$.
\item Perform two dimensional Fourier transform (real to complex)
of the tapered $T.I$.
\item Evaluate the convolution (equation \ref{EQN:WGCFFT}) for each sample point
to obtain the predicted visibility.
\end{enumerate}

To evaluate the dirty (non-deconvolved) image:

\begin{enumerate}
\item For each sample visibility, evaluate the convolution 
for a gridded $(u,v,w=0)$ plane.
\item Perform two dimensional inverse Fourier transform (complex to
real) to obtain the tapered $T.I^D$ (where $I^D$ denotes the dirty image).
\item Divide out the image plane tapering function $T$ to obtain
the dirty image.
\end{enumerate}

Numerical integration is required to find $\tilde{G_T}$. We use 4x
padding in the image plane, equal spacing of $\sqrt{w}$ planes, and
truncation of the aggregate convolution function (at 0.1\%). These
values have been determined empirically from the requirement that
aliasing is suppressed at a dynamic range of $10^4$ or better. For
better performance, it will be worthwhile to design a 2D FIR filter to
approximate $\tilde{G_T}$ more exactly.

In figure (3), we show the image plane function and Fourier transform
for $w=0, w_{max}/2, w_{max}$ for a typical observation with the
VLA. For $w=0$, the convolution function is simply the usual
spheroidal function. For $w$ increasing, the area of the $(u,v)$ plane
affected by the projection of any point increases approximately as
$\sqrt{w}$.

\begin{center}
\begin{figure*}[htb]
\label{FIG:WPLANES}
\begin{center}
\includegraphics[width=6cm]{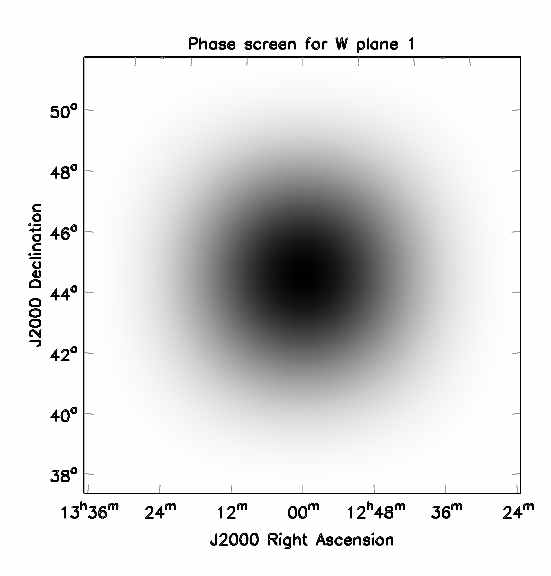}
\includegraphics[width=6cm]{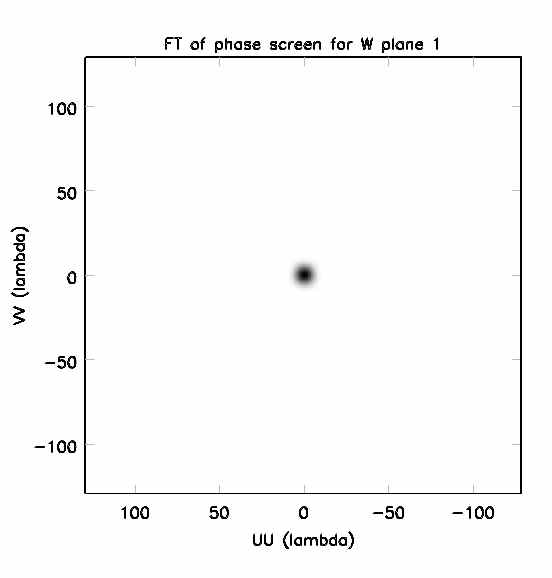}\\
\includegraphics[width=6cm]{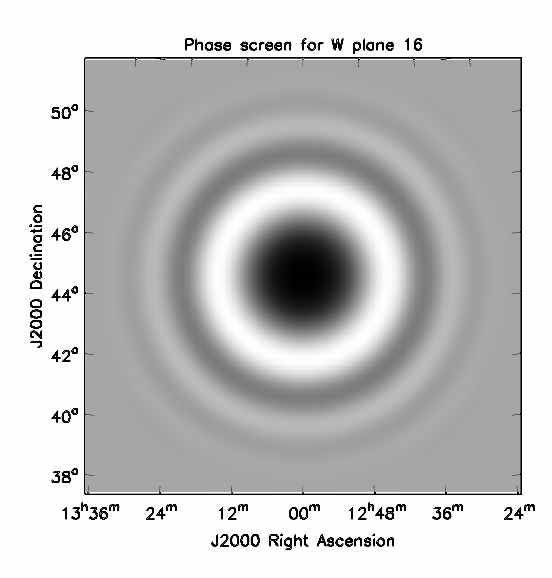}
\includegraphics[width=6cm]{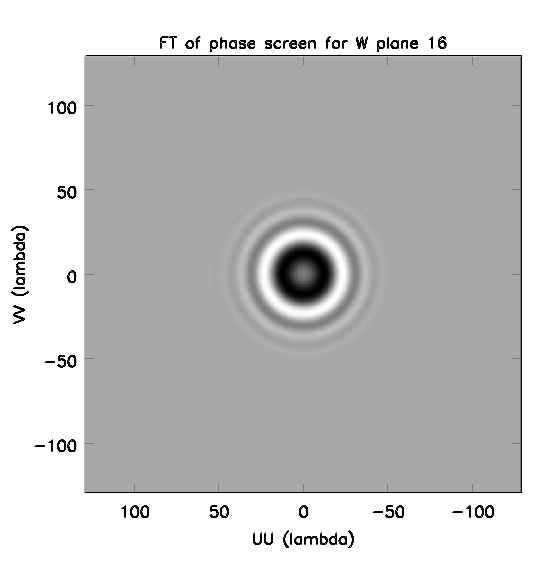}\\
\includegraphics[width=6cm]{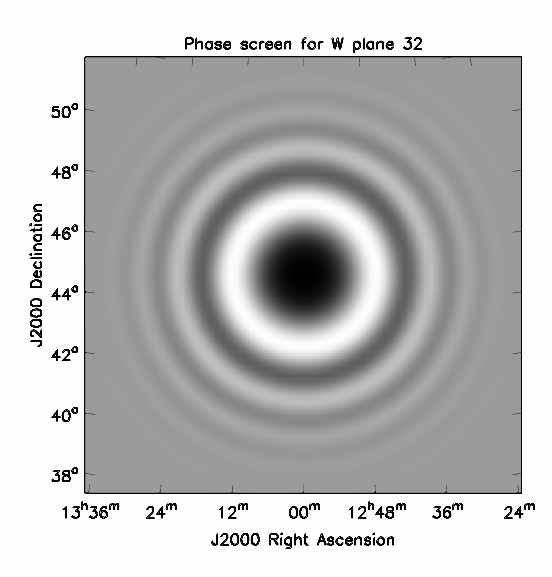}
\includegraphics[width=6cm]{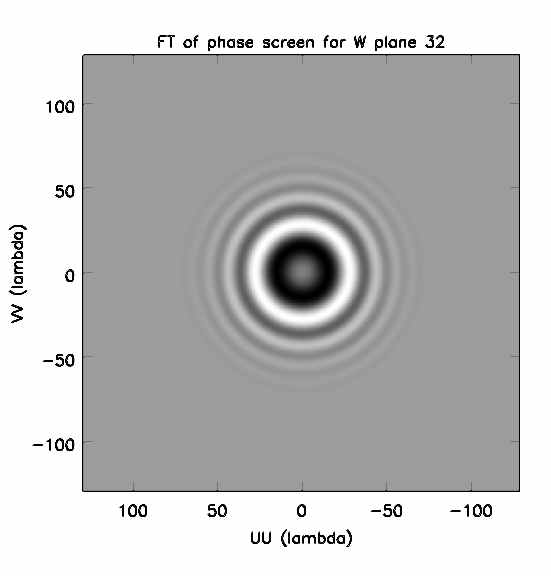}\\

\end{center}
\caption{Image plane function and Fourier transform for
(top) $w=0$, (middle) $w=w_{max}/2$, (bottom) $w=w_{max}$. The range of
brightness is -1 to +1.}
\end{figure*}
\end{center}

For the image plane tapering function $T$, we use a spheroidal
gridding function. The usual spheroidal gridding function has support
9 by 9 pixels in $(u,v)$. The support of $\tilde{G_T}$ grows with both $w$ and
the field of view, typically up to a largest value of about 70 by 70.
The maximum memory required per plane is a few MB, for a total of up to
1GB - an amount which is common in desktop computers only in the last 
few years. The work involved in convolving with this function increases 
in direct
proportion to the extent in $(u,v)$. The average value of $|w|$
gridded is close to $w_{max}/2$, so the typical increase in gridding
costs is about a factor of 10-20. While increasing the spread of $\tilde{G_T}$
in $(u,v)$ directly leads to more computing, the extra cost incurred
by using more planes in $w$ is relatively minor - all that happens is
an essentially negligible increase in the memory access time required
to get the relevant part of the convolution function. 

Tabulation of $\tilde{G_T}$ in $w$ leads to aliases at the scale size of
$1/\Delta w$. Placing these aliases outside the field of view requires
that $w$ be sampled with similar precision to $(u,v)$. Tabulating
planes evenly in $\sqrt{w}$ reduces this effect to a tolerable
level. With this in place, the remaining errors are due to the
incorrect value of the convolution function being used. Since this
aliasing error is localized in $\sqrt{w}$, the resulting error is not coherent
in the image plane. This is in contrast to the facet approaches
where the prediction error is necessarily worse on the longest
baselines, and so the resulting error patterns are quite coherent in
the image.

\subsection{Computational load}

Most of the computational load lies in the gridding/degridding step
which is directly determined by the number of visibility samples and
the size of the convolution kernel.  Taking the field of view to be
$\lambda/D$, we find that the number of pixels needed in the
convolution kernel along each of the $u$ and $v$ axes goes as
$B\lambda /D^2$.

The costs for facet based gridding go as the total number of facets ({\em i.e.} the product of the
number along each of the two spatial axes). We then have that the costs per sample go as:

\begin{eqnarray}
\label{EQN:COSTS}
t_{facets} &=& N^2_{facets}.N^2_{GCF points}.t_{single}\\
t_{w project} &=& 2 (N^2_{w project}+N^2_{GCF points}).t_{single}
\end{eqnarray}

In this equation, $N_{GCF points}$ is the support of the normal
gridding convolution function in one axis (typically 9), and
$t_{single}$ is the time to grid a single sample to a single grid
point. $N_{facets}$ (the number of facets in one axis) and $N_{w
project}$ (the typical size of the $\tilde{G_T}$ gridding function)
are both proportional to $B/(\lambda D^2)$ but with different
proportionality constants.  We have assumed that the sizes of the
normal gridding convolution function and $\tilde{G}$ add in
quadrature. Note also that $\tilde{G_T}$ is necessarily complex. If we
take $N_{facets}$ and $N_{w project}$ to be roughly equal, then for
large fields of view, the ratio of these times is roughly the total
number of points in the {\em normal} gridding convolution
function. Since the normal gridding convolution function is typically
7 by 7 or 9 by 9, the {\em asymptotic} speedup is between 25 and
50. Allowing for different proportionality constants, we could
conservatively expect at least an order of magnitude speed advantage
for W-projection. The gains for less severe non-coplanarity will be
smaller. In the next section, we examine the relative performance for
simulated data.

At first sight, it is curious that W-projection should be faster than
the facet approaches. However, we can see from the analysis just given
that the discrepancy comes from the relative inefficiency of using a
broad convolution function in the standard FT approach. If box-car
convolution were to be used in facet based imaging, the computing
costs would be roughly the same for the two approaches.

\subsection{Implementation} We have implemented the various algorithms 
(standard FT, uvw-facets, and W-projection) in \aipspp. Most of the
code is C++ using \aipspp\space  libraries, but inner loops for gridding and
Fourier transformation are written in hand-optimized Fortran. Possible
minor cycle deconvolution steps being standard CLEAN, a multi-scale
CLEAN algorithm \citep{BhatnagarCornwell2004}, or a Maximum Entropy
algorithm \citep{CornwellEvans1985}.

\section{Simulated data}

Non-isoplanatism is most troublesome for fields full of emission, and
the non-coplanar baselines effect is greater for longer
wavelengths. Accordingly, we simulated a low frequency observation of
a typically full field: a 74MHz VLA C-configuration full synthesis on
a field at right ascension 12h56m57.18, declination 47d20m20.801. The
data consisted of 505440 visibility records, each of 8 spectral
channels. The sources were generated by taking the sixty six  Westerbork Northern Sky Survey (WENSS) \citep{WENSS97} 
sources brighter than 2Jy within 12 degrees of the specified
center. The sources were scaled to 74MHz by a spectral index of -0.7,
and then multiplied by a simple model of the 74MHz antenna primary
beam. The data corresponding to these scaled sources were calculated
using analytical transforms, and should thus be fully accurate to
machine precision. The brightest source has strength 47.8Jy and has
been chosen to be at the field center. The data generated is somewhat
realistic for a randomly chosen 74MHz field observed with the VLA, the
major deficiency being the lack of weak sources.

The processing was performed on a dual Xeon 3.06GHz machine with
per-processor cache 512k and 3GB of memory. AIPS++ was compiled
using the GNU suite of compilers with optimization flags set
to O2.

For this type of confused field observed with the VLA, deconvolution is necessary.
All images were cleaned with 20,000 iterations at a loop gain of
0.1. Both uvw-space facet and W-projection algorithms used in-memory
gridding and Fourier transform.  The image size was 1536 by 1536
pixels of 60arcsec. The maximum support of the gridding function
$\tilde{G_T}$ is 220 pixels (full width). We use a gridding function
of size 512 by 512 complex pixels for each plane so the memory required for
storage of the gridding function is 2MB per plane.

The results are shown in Table I and displayed in Figure 4. The key points are:

\begin{itemize}
\item For the specific case tested here, W-projection is about 20 times 
slower than standard two dimensional Fourier transformation. We may
understand this as follows: The standard gridding uses support of 9 by
9 pixels. The typical size of the convolution region in W-projection
is about 30 by 30 pixels so the increased load in gridding for W-projection should be about (2*30*30/9*9) times bigger or a factor of
24. This rough agreement indicates that the W-projection costs are in
proportion to the increase in multiplications in the innermost loop.
\item For the same dynamic range, W-projection is about an order of
magnitude faster than uvw-space facets.
\item For W-projection, the aliasing performance and initialization 
time both grow linearly with the number of planes, whereas the
residual calculation time is independent of the number of planes (in
fact a slight decrease is seen as the number of planes increases).
\item For uvw-space facets, the aliasing performance and the residual
calculation times both grow linearly with the number of facets ({\em
i.e.} the inverse square of Fresnel number).
\end{itemize}

\begin{table}
\caption{Table I: Performance of various methods on simulated 74MHz C 
configuration observations. The first column shows the approach used,
the second is a robust estimate of the dynamic range after cleaning
(DR1=peak/(median absolute deviation from the median)), the third is
the dynamic range on a bright source (DR2=peak/abs(max near
negative)), the fourth is the time in seconds to initialize the
convolution functions, the fifth is the time taken to calculate the
residual image (involving a transform in both directions), and the
last is the total time to deconvolve.}
\label{TAB:4BAND}
\begin{center}
\begin{tabular}{lccccc}
  \hline
  Method & DR1 & DR2 & Calc G & Calc res & Total\\ \hline
  \hline
  FT & 1570 & 2.8 & - & 27 & 216 \\
  uvw-facets (3 by 3) & 2550 & 3.6 & - & 204 & 784\\
  uvw-facets (9 by 9) & 12580 & 40 & - & 1348 & 30488 \\
  w-proj (8) & 2692 & 17& 9 & 671 & 2748 \\ 
  w-proj (16) & 4665 & 53 & 18 & 661 & 2688\\
  w-proj (32) & 9066 & 268 & 35 & 647 & 3985 \\
  w-proj (64) & 18198 & 524 & 83 & 607 & 3161 \\
  w-proj (128) & 33370 & 856 & 103 & 532 &  3857 \\
  w-proj (256) & 50888 & 984 & 231 & 507 & 5419\\
\hline
\end{tabular}
\end{center}
\end{table}

\begin{center}
\begin{figure*}[htb]
\label{FIG:PERF}
\begin{center}
\includegraphics[width=15cm]{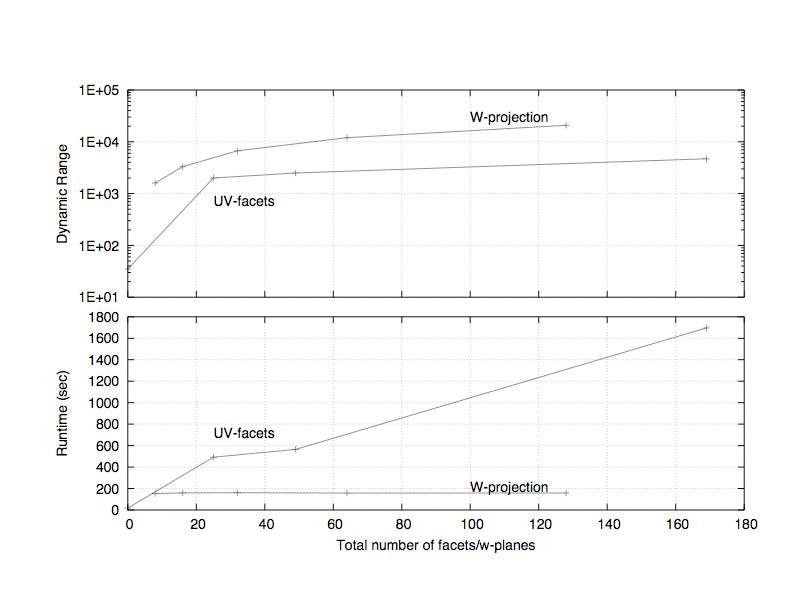}
\end{center}

\caption{Performance of the two methods: uv-facets and W-projection.
  Lower panel shows the actual runtime (smaller is better) as a
  function of total number of uv-facets $N^2_{facets}$ for the
  uv-facets algorithm and as a function of total number of w-planes
  for the W-projection algorithm.  Top panel shows the achieved
  dynamic range (larger is better) in the simulations as a function of
  the number of uv-facets/w-planes.  For a given number of
  facets/w-planes, the runtime and the dynamic range both are in
  favour of W-projection algorithm.  }
\end{figure*} 
\end{center}

In figure (5), we compare the restored images obtained from standard
FT, uvw-space facets (9 by 9), and W-projection (256). A few
comments:

\begin{itemize}
\item The standard FT image shows severe distortion of sources 
away from the phase center. For the uvw-facets approach, the
distortion is reduced but is still present for sources away from a
facet center (as shown by DR2).
\item The errors in W-projection image are much more isotropic (as
evidenced by comparison of the DR1 and DR2 numbers).
\item The computing time differential between uvw-facets and W-projection 
in residual calculation is about a factor of thirty, and the error
performance about 3, both in favor of W-projection.
\item The uvw facets algorithm has a relative advantage early
in iteration because it need not Fourier transform (to the visibility
plane) any facets empty of emission. This partially accounts for the
difference between the factor of thirty for residual calculation and
six overall. In addition, the scaling of the Clean minor cycle
slightly favors the uvw facets algorithm.
\item Extrapolating the uvw-space facets behavior, we find that to 
obtain the same dynamic range as the best W-projection image, about 27
by 27 facets would be required, at a cost of about 100 compared to w
projection.
\end{itemize}

\begin{center}
\begin{figure*}[htp]
\centering
\label{FIG:COMPARISON}
\subfigure[standard Fourier transform]{
\includegraphics[width=6cm]{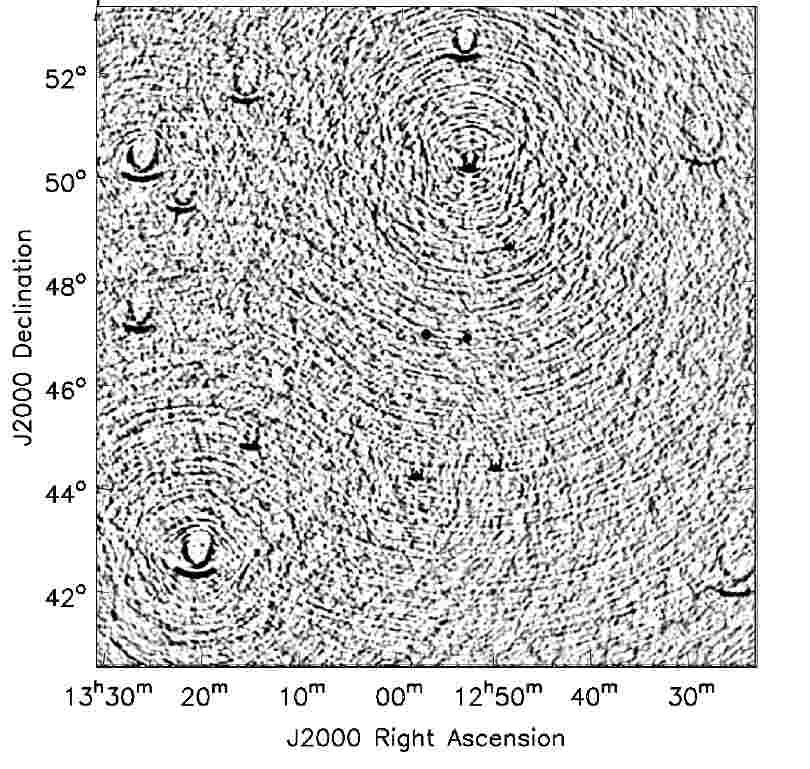}
\hspace{0.01cm}
\vspace{0.01cm}
}
\hspace{0.01cm}
\vspace{0.01cm}
\subfigure[uvw-space facets (9 x 9)]{
\includegraphics[width=6cm]{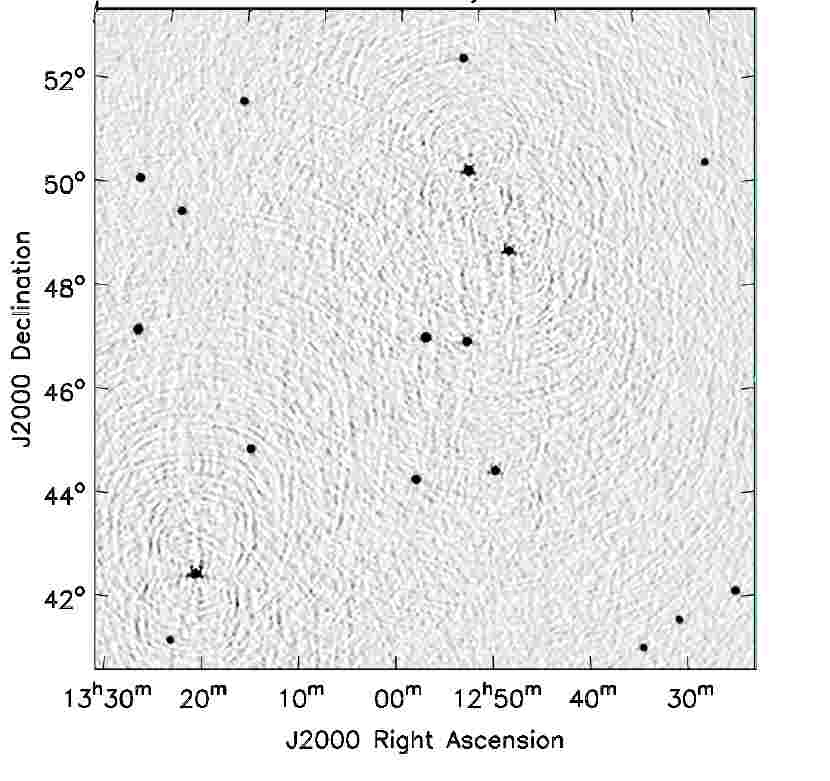}
\vspace{0.01cm}
\hspace{0.01cm}
}
\hspace{0.01cm}
\vspace{0.01cm}
\subfigure[W-projection
  (128 ${\tilde G_T}$ planes)]{
\includegraphics[width=6cm]{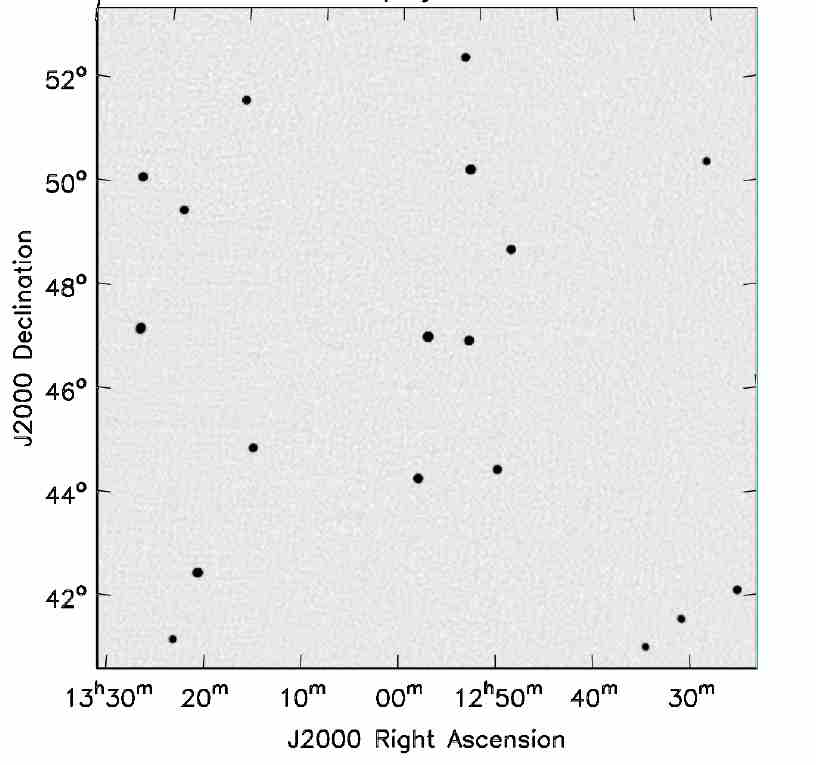}
\vspace{0.01cm}
\hspace{0.01cm}

}

\caption{Clean images for 74MHz simulation.  The brightness range is -5 to +50 mJy/beam, and the peak
  brightness should be 47.2Jy. The peak sidelobes around the brightest
  sources in the uvw-space facets image are about 0.3\%.  Calculation
  of these images took 784s, 30488s, and 5419s respectively.}

\end{figure*}
\end{center}

\section{A wide-field image of the SN1006 at 1.4GHz}

As an example of the application of W-projection to real data, we show an image of
the supernova remnant SN1006 observed at 1.4GHz. The data are from  all the four  configurations of the Very Large Array (VLA)  and Green Bank Telescope (GBT). Details
of the data reduction and scientific interpretation are given elsewhere \cite{Dyeretal2005}.
The GBT image was deconvolved using a Maximum Entropy algorithm, and this
image was then used as a starting point for a multi-scale CLEAN algorithm \cite{BhatnagarCornwell2004}.
Our W-projection  approach was used for the major cycle of the CLEAN algorithm.
An image plane facet based algorithm would require at least 10 by 10 facets, thus crisscrossing
the extended emission. For W-projection, we used 256 planes. 

\begin{center}
\begin{figure*}[htb]
\label{FIG:GC}
\begin{center}
\includegraphics[width=9cm]{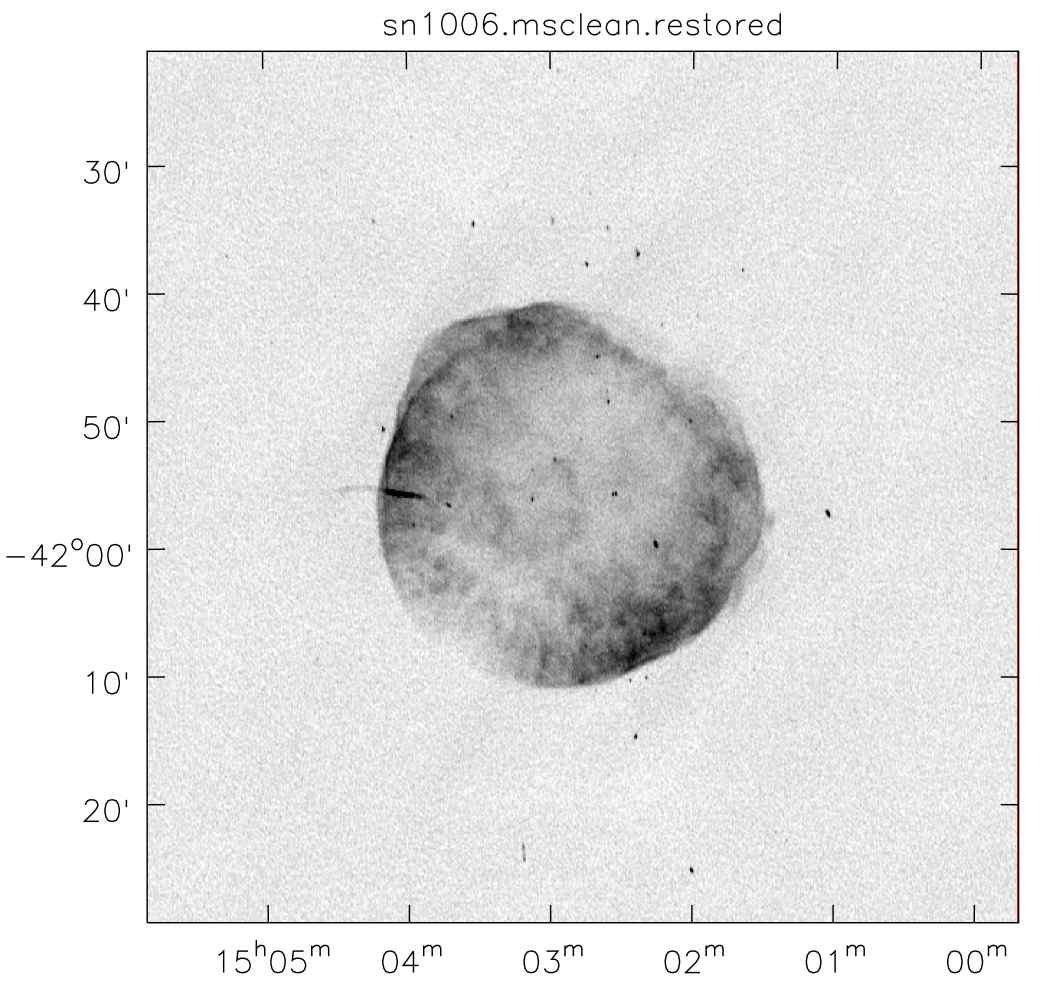}
\end{center}

\caption{Multi-scale CLEAN image of the SN1006 made using W-projection (256 planes) 
and a multi-scale CLEAN algorithm. A facet-based algorithm would have required
approximately 10 by 10 facets.}

\end{figure*}
\end{center}

\section{Implications}

The results in this paper have a number of implications for the design
and operation of wide-field imaging radio synthesis telescopes.

First, the meaning of redundancy in wide-field imaging is much more
restrictive than often thought. To get the same visibility, one must
measure at exactly the same $(u,v,w)$ - having identical $(u,v)$ is
not adequate. This means that redundant self-calibration or
data-editing is limited in applicability for wide-field imaging.

Second, calculating images for wide field of view will be less
demanding than expected. Our algorithm is substantially faster, and
scales much more slowly with desired precision than the facet based
algorithms. As shown in the simulations, the precision of the facet
approaches scales roughly as the square root of the computational
time, whereas that for W-projection is roughly independent of the
computational time, the cost mainly being in memory.

\citet{PerleyClark2003} derived scaling relationships for the
computing costs associated with wide-field imaging as a function of antenna
diameter and baseline length. They based their analysis on faceted
algorithms. As shown above, the computational costs of W-projection
can be less by about an order of magnitude. This affects only the
coefficients in the Perley-Clark cost equation, not the form.
However, an order of magnitude is a very substantial gain, equivalent
to about 5 years of Moore's Law growth. \citet{Cornwell2005} has evaluated
computing costs for the Square Kilometer Array and Expanded Very Large
Array in light of the W-projection algorithm.

The cost of convolution with $\tilde{G_T}$ can get to be quite large. We show
representative values for the size of $\tilde{G_T}$ in Table II. The very
large values found for few hundred kilometer baselines at meter
wavelengths may require another shift in algorithm design, perhaps to
a hybrid W-projection/uvw facets approach. Whether this is so depends
upon the architecture of the machines on which such processing will be
performed. It is clear that the memory requirements currently prohibit
W-projection on a single processor computer for the very worst cases, unless
some symmetry properties are exploited.

\begin{table}
\caption{Table II: Support of $\tilde{G_T}$ (half-width in pixels)
as a function of maximum baseline length and wavelength. These have
been scaled from values found in the simulation described above. For the
diameter of the VLA antennas, 25m, these  numbers are within a few
percent of the Fresnel radius $r_F$ in meters.}

\begin{center}
\label{TAB:SUPPORT}
\begin{tabular}{lccccc}
  \hline
  Wavelength& & 4m & 1m & 0.2m & 0.06m\\ \hline
  Baseline (m) & Image size & & & &\\ \hline
  \hline
  1000 & 240 & 58 & 29 & 13 & 7 \\
  3500 & 840 & 109 & 54 & 25 & 13 \\
  10000 & 2400 & 184 & 92 & 42 & 23 \\
  35000 & 8400 & 344 & 172 & 79 & 42 \\
  100000 & 24000 & 582 & 291 & 133 & 71 \\
  350000 & 84000 & 1089 & 544 & 249 & 133 \\
\hline
\end{tabular}
\end{center}
\end{table}

This naturally brings up the question of how best to parallelize the
W-projection algorithm. Parallelization of the image- and uvw
space-facets algorithms has been demonstrated \citep{Cornwell1993,
Golapetal2001}.  The facet-based algorithms are parallelized by
delegating the residual calculation for each facet to a separate
processor. Each processor must access every record of a visibility
data set and communicate back the cleaned residual image for a single
facet. Golap {\em et al.} \citep{Golapetal2001} demonstrated
reasonable scaling for imaging of 15 by 15 facets for up to 32
processors, linear up to 16 and flattening slightly at 32. This
flattening is thought to be due to I/O blocking, and so parallel I/O
may be necessary. For W-projection, the natural partitions are the
separate planes in $\sqrt{w}$. In this scenario, the gridding for each
$\sqrt{w}$ plane is delegated to a separate processor which need only
store the visibility data for that plane and at the end of gridding,
the separate image grids are simply added, either before or after
Fourier transform. This has the side-benefit of reducing the memory
cost per processor to an affordable level. Each processor has to read
only a small sub-set of the records in a visibility data set, and must
communicate back a copy of the entire residual image. Parallelization
of the minor-cycle deconvolution depends on the algorithm used.

\section{Summary}

We have demonstrated a new algorithm for correcting the non-coplanar
baselines effect. This has superior performance in both speed and
error control, at the cost of greater memory usage. These advantages
grow more substantial as the non-coplanarity grows more severe.
It does not make it necessary to use only pixel based conventional CLEAN 
deconvolution step and thus is well-suited to wide-field imaging
of very extended emission (for example, the Galactic plane) where
multi-scale or Maximum Entropy methods are required.

The facets algorithms remain roughly competitive with W-projection
only at low dynamic range, for example that occuring in VLA
observations at 74MHz.  At high sensitivities, W-projection will be
decisively superior.  Hence use of this algorithm (and similar
algorithms for other problems) will have a substantial impact on the
predicted computing costs for new radio synthesis arrays such as the
EVLA, LOFAR, and SKA.

\bibliographystyle{aa}
\bibliography{wprojection}

\begin{thebibliography}{20}
\expandafter\ifx\csname natexlab\endcsname\relax\def\natexlab#1{#1}\fi


\bibitem{BhatnagarCornwell2004}
{Bhatnagar}, S. \& {Cornwell}, T., Scale Sensitive Deconvolution of Interferometric Images,  2004, \aap, 426, 747-754 

\bibitem[{{Bracewell}(1983)}]{Bracewell1983}
{Bracewell}, R. 1983, Inversion of Non-Planar Visibilities. In Measurement and Processing for
  Indirect Imaging. Proceedings of an International Symposium held in Sydney,
  Australia, August 30-September 2, 1983. Editor, J.A. Roberts; Publisher,
  Cambridge University Press, Cambridge, England, New York, NY, 1984. P.177

\bibitem[{Brouw(1969)}]{Brouw1969}
Brouw, W. 1969, Data Processing for the Westerbork Synthesis Radio Telescope
  (Ph.D. thesis, Univ. of Leiden)

\bibitem[{{Clark}(1980)}]{Clark1980}
{Clark}, B., An efficient implementation of the algorithm 'CLEAN', 1980, \aap, 89, 377

\bibitem[{{Clark}(1980)}]{Clark1973}
{Clark}, B., Curvature of the Sky, 1973, VLA Scientific Memo 107 (National Radio Astronomy Observatory, http://www.nrao.edu)

\bibitem[{{Cornwell}(1988)}]{Cornwell1988}
{Cornwell}, T., Radio-interferometric imaging of very large objects, 1988, \aap, 202, 316

\bibitem[{Cornwell(1993)}]{Cornwell1993}
Cornwell, T., Improvements in Wide-Field Imaging, 1993, VLA Scientific  Memo 164 (National Radio Astronomy Observatory,
  http://www.nrao.edu)

\bibitem[{Cornwell(2004)}]{Cornwell2005}
Cornwell, T.,  Ska and Evla Computing Costs for Wide Field Imaging, 2004, Experimental Astronomy 17:11, 329-343, 6/2004

\bibitem[{{Cornwell} \& {Evans}(1985)}]{CornwellEvans1985}
{Cornwell}, T. \& {Evans}, K., A simple maximum entropy deconvolution algorithm 1985, \aap, 143, 77


\bibitem[{{Cornwell} {et~al.}(1995){Conwell}, {Briggs}, \&
  {Holdaway}}]{SDE1995}
{Cornwell}, T., {Briggs}, D., \& {Holdaway}, M. 1995, {User Guide to SDE} (NRAO, Socorro)


\bibitem[{{Cornwell} {et~al.}(1993){Cornwell}, {Holdaway}, \&
  {Uson}}]{Cornwelletal1993}
{Cornwell}, T., {Holdaway}, M., \& {Uson}, J., Radio-interferometric imaging of very large objects: implications for array design, 1993, \aap, 271, 697

\bibitem[{{Cornwell} \& {Perley}(1992)}]{CornwellPerley1992}
{Cornwell}, T. \& {Perley}, R., 
	Radio-interferometric imaging of very large fields - The problem of non-coplanar arrays, 1992, \aap, 261, 353

\bibitem[{Dyer(2005)}]{Dyeretal2005}
Dyer, K., {\em et al.}, High-resolution Flux-accurate Radio Images of SN1006, 2005, Bulletin of the American Astronomical Society, Vol. 37, p.1437

\bibitem[{{Ekers}(2003)}]{Ekers2003}
{Ekers}, R., Square Kilometre Array (SKA), 2003, in ASP Conf. Ser. 295: Astronomical Data Analysis Software
  and Systems XII, 125--+

\bibitem[{{Frater} \& {Docherty}(1980)}]{FraterDocherty1980}
{Frater}, R. \& {Docherty}, I., On the reduction of three dimensional interferometer measurements, 1980, \aap, 84, 75

\bibitem[{{Golap} {et~al.}(2001){Golap}, {Kemball}, {Cornwell}, \&
  {Young}}]{Golapetal2001}
{Golap}, K., {Kemball}, A., {Cornwell}, T., \& {Young}, W., 
	Parallelization of Widefield Imaging in AIPS++, 2001, in ASP Conf.
  Ser. 238: Astronomical Data Analysis Software and Systems X, 408

\bibitem[{Goodman(2002)}]{Goodman2002}
Goodman, J. 2002, Introduction to Fourier Optics (McGraw Hill)

\bibitem[{{Perley}(2000)}]{Perley2000}
{Perley}, R., The Very Large Array Expansion Project, 2000, in Proc. SPIE Vol. 4015, p. 2-7, Radio Telescopes, Harvey R.
  Butcher; Ed., 2--7

\bibitem[{Perley \& Clark(2003)}]{PerleyClark2003}
Perley, R. \& Clark, B., Scaling Relations for Interferometric Post-Processing, 2003, EVLA memo 63 (National Radio Astronomy
  Observatory, http://www.nrao.edu)

\bibitem[{{Sault} {et~al.}(1999){Sault}, {Bock}, \& {Duncan}}]{Saultetal1999}
{Sault}, R., {Bock}, D.-J., \& {Duncan}, A., Polarimetric imaging of large fields in radio astronomy, 1999, \aap, 139, 387

\bibitem[{{Taylor} {et~al.}(1999){Taylor}, {Carilli}, \&
  {Perley}}]{Tayloretal1999}
{Taylor}, G., {Carilli}, C., \& {Perley}, R., eds. 1999, {Synthesis Imaging in
  Radio Astronomy II}

\bibitem[{{Perley} {et~al.}(1989){Perley}, {Schwab}, \&
  {Bridle}}]{Perleyetal1989}
{Perley}, R., {Schwab}, F. R., \& {Bridle}, A.H., eds. 1999, {Synthesis Imaging in
  Radio Astronomy}

\bibitem[{{Thompson} {et~al.}(2001){Thompson}, {Moran}, \&
  {Swenson}}]{Thompsonetal2001}
{Thompson}, A., {Moran}, J., \& {Swenson}, G. 2001, {Interferometry and
  synthesis in radio astronomy} (Wiley, New York)


\bibitem[{{Rengelink} {et~al.}(1997){Rengelink}, {Tang}, {de Bruyn}, {Miley}, {Bremer}, {Rottgerring} \&
  {Bremer}}]{WENSS97}
{Rengelink}, R. B., {Tang}, Y., {de Bruyn}, A. G., {Miley}, G. K, {Bremer}, M. N., {Rottgering}, H. J. A. \& {Bremer}, M. A. R.,The Westerbork Northern Sky Survey (WENSS),  1997, Astron. Astrophys. Suppl., 124, 259-280

\end{thebibliography}

\section* {Acknowledgement}

We thank the \aipspp\space Project team for providing a superb software
environment in which to do this development, resulting in the
happy situation where most of the time in this project went to thinking
and experimentation rather than the coding. TJC thanks Wim Brouw for
his patience in explaining simple geometry via email, and David King
for many stimulating and insightful questions. We all thank the
Fridays-on-Thursdays group in Socorro for many hours of conversation
relevant to imaging. Kristy Dyer kindly allowed us to use her SN1006 data.

\section*{Appendix A: General structure of imaging algorithms}

The scientific impact of radio interferometry has been heavily
influenced by the advent of deconvolution and self-calibration
algorithms, starting with Clean in the mid-seventies 
\citep[see][]{Thompsonetal2001}.  The introduction of the minor/major
cycle approach (such as the Clark CLEAN algorithm and the
Cotton-Schwab CLEAN) has aided the generation of new deconvolution
algorithms. In such algorithms, the deconvolution is split into two
cycles, the minor cycle using an approximate PSF for computational
speed, and the major cycle using a full calculation for accuracy. Much
of imaging then reduces to two steps:

\begin{itemize}
\item Finding and solving an approximate convolution problem for the
minor cycle.
\item Performing the major cycle with computational efficiency and
high numerical accuracy.
\end{itemize}

First, we should clarify a few definitions. Nearly all of the pure
imaging problems are linear in the pixel values (unlike the calibration
problem, which is non-linear in the unknown antenna gains). We can
therefore write a linear equation:

\begin{equation}
\label{EQN:ME}
D = A I + N
\end{equation}

where $D$, $I$ and $N$ are vectors for data, image, and noise, and $A$
is the (non-square) observation matrix. In the usual case of simple
radio interferometry, the elements of $A$ are the cosines and sines of
the Fourier transform. The observation matrix is then usually singular
and cannot be simply inverted. Instead, it is
generally the case that non-linear iterative methods are used to solve
this linear equation. 

The normal equations arising from least squares minimization of the
image pixels are:

\begin{equation}
\label{EQN:NORMAL}
A^T D = A^T A I
\end{equation}

For the simple case where $A$ represents a Fourier transform (or
discrete sum), $A^T D$ is the dirty image, and $A^T A I$ is the true
sky $I$ convolved with the dirty beam $A^T A$.  (Note that we have
left out the covariance matrix of the noise for simplicity, though in
practice, this must be taken into account).

Interpreting the deficit as a residual image, we have:

\begin{equation}
\label{EQN:RESIDUAL}
I^R = A^T (D - A I)
\end{equation}

Calculating $I^R$ for a given model image $I$ constitutes the major
cycle. This consists of two steps - first the calculation of the
residual data $D-AI$, and second the calculation of the residual image
$A^T(D-AI)$. The first step must be done with high accuracy but the
second need only be done approximately. Let $B$ be an approximation
to $A^T$; we then find it convenient to work with the approximate
residual image:

\begin{equation}
\label{EQN:APPROXRESIDUAL}
I^{AR} = B (D - A I)
\end{equation}

The minor cycle consists of finding an update to the image $I$ given
either the residual image $I^R$ or the approximation $I^{AR}$. In some
circumstances, the minor cycle need not be done with high accuracy
because any errors will be corrected in the major cycle. Since most of
our most efficient algorithms are for shift-invariant PSFs, it pays to
try to find an approximate shift-invariant convolution to be solved in
the inner loop. For example, for a homogeneous mosaic, the linear
mosaic of the individual dirty (residual) images can be approximated
as a convolution of the true image with an approximate PSF \citep{Cornwelletal1993}.
As another example, the nominally
shift-variant PSF encountered in the non-coplanar baselines effect
can be often be approximated by a shift-invariant
PSF for the minor cycle.

The computational costs of deconvolution thus split into two
parts:
\begin{itemize}
\item Calculating the normal equations: specifically the predicted
data $AI$, and the residual image $A^T(D-AI)$.
\item Updating the model using the residual image $A^T(D-AI)$.
\end{itemize}

\end{document}